\documentstyle[12pt]{article}
\newcommand{\be}[1]{\begin{equation}\label{#1}}
\newcommand{\ba}[1]{\begin{eqnarray}\label{#1}}
\newcommand{\ee}{\end{equation}}
\newcommand{\ea}{\end{eqnarray}}
\newcommand{\non}{\nonumber\\\rule{0pt}{30pt}}
\newcommand{\nona}[1]{\nonumber\\\rule{0pt}{#1pt}}

\newcommand{\dis}{\displaystyle}
\newcommand{\eq}[1]{(\ref{#1})}
\newcommand{\vac}{|0\rangle}
\newcommand{\dvac}{\langle0|}
\newlength{\HFPP}       \HFPP5.4mm
\makeatletter
\@addtoreset{equation}{section}
\makeatother

\newtheorem{thm}{Theorem}[section]
\begin{document}
\begin{center}
\begin{Large}
{\bf The form factors in the finite volume}
\end{Large}

\vspace{20pt}
\begin{large}
V. E. Korepin\raisebox{2mm}{{\scriptsize $\dagger$}}
		    and
N. A. Slavnov\raisebox{2mm}{{\scriptsize $\ddagger$}}
\end{large}

\vspace{40pt}

~\raisebox{2mm}{{\scriptsize $\dagger$}}
{\it ITP, SUNY at Stony Brook, NY 11794-3840, USA.}\\

\vspace{20pt}

~\raisebox{2mm}{{\scriptsize $\ddagger$}}
{\it Steklov Mathematical Institute,
Gubkina 8, Moscow 117966, Russia.}\\

\vspace{32pt}
Abstract
\vspace{16pt}

\noindent\parbox{12cm}{\small The form factors of integrable models 
in finite volume are studied. We construct the explicite 
representations for the form factors in terms of determinants. }

\end{center}

\vspace{32pt}
\section{Introduction}

In this  paper we study form factors of quantum integrable 
models in finite volume. The possibility to obtain exact 
non-pertubative results in this field is of great importance. 
The Quantum Inverse Scattering Method created by L.~D.~Faddeev and 
his school \cite{1,2,3} allows one to formulate general methods of 
solving this problem. Namely, in order to obtain an explicit 
representation for a form factor of an operator ${\cal O}$, it is 
sufficient to embed this operator into the algebra
\be{0RTT-} 
R(\lambda,\mu)\biggl(T(\lambda)\otimes T(\mu)\biggr) 
=\biggl(T(\mu)\otimes T(\lambda)\biggr)R(\lambda,\mu).
\ee
After this is done, the problem is reduced to the standard Algebraic 
Bethe Ansatz procedures.

We would like to mention that our approach is drastically different 
from the ones, used in relativistic models \cite{4}--\cite{N}.  We 
consider form factors between the states constructed on the bare 
vacuum, instead of the physical one. We also do not use the axioms, 
describing analytical properties of form factors. All our 
calculations are based on the algebra \eq{0RTT-} only. In spite of 
this approach meets serious combinatorial difficulties, nevertheless
it provides us with explicit determinant representations for form 
factors. 

In the present paper we consider Quantum Nonlinear Schr\"odinger 
equation (QNLS) and $XXZ$ ($XXX$) Heisenberg chains of spin $1/2$.
The QNLS model in a finite volume $L$ with periodically
boundary conditions is described by the Hamiltonian
\be{0HamQNLS}
H_{\mbox{\scriptsize QNLS}}=\int_{-L/2}^{L/2}\,dx\left(
\partial_x\Psi^\dagger\partial_x\Psi+ c\Psi^\dagger
\Psi^\dagger\Psi\Psi\right),
\ee
where $\Psi(x,t)$ and $\Psi^\dagger(x,t)$ are canonical Bose
fields
\be{0Bosefield}
[\Psi(x,t),\Psi^\dagger(y,t)]=\delta(x-y),\qquad
\Psi\vac=0,\qquad\dvac\Psi^\dagger=0,
\ee
and  $c$ is the coupling constant.

The Hamiltonian of the $XXZ$ Heisenberg chain with $M$ sites is
\be{0HamXXZ}
H_{XXZ}=-\sum_{m=1}^{M}\left(
\sigma_-^{(m)}\sigma_+^{(m+1)}+\sigma_+^{(m)}\sigma_-^{(m+1)}
+\Delta(\sigma_z^{(m)}\sigma_z^{(m+1)}-1)\right).
\ee
Here $\sigma_a^{(m)}$ are local spin operators (Pauli matrices),
associated to the $m$-th site, $\sigma_\pm=\frac{1}{2}
(\sigma_x\pm i\sigma_y)$, \ $\sigma_a^{(M+1)}=\sigma_a^{(1)}$.
Parameter $\Delta$ defines the anisotropy of the model. Particular
choice $\Delta=1$ corresponds to the $XXX$ chain.

Let us describe now the content of the paper. In the section 2 we 
recall the basic definitions of the Algebraic Bethe Ansatz. In the 
section 3 the determinant representation for the form factor of the 
local field $\Psi$ of QNLS is obtained. In the section 4 we consider
different representations for form factors of QNLS and prove their
equivalency. The last section is devoted to the form factors
of Heisenberg chains.

\section{Algebraic Bethe ansatz}

We consider integrable models, which can be solved via the 
Algebraic Bethe Ansatz.  Let the monodromy matrix is 
\be{0mon}
T(\lambda)=\left(
\begin{array}{cc} A(\lambda)&B(\lambda)\\C(\lambda)&D(\lambda)
\end{array}\right),
\ee
and the highest vector $\vac$ (pseudovacuum) of this matrix
has the following properties:
\be{0vac} 
A(\lambda)\vac=a(\lambda)\vac,\quad 
D(\lambda)\vac=d(\lambda)\vac,\quad 
C(\lambda)\vac=0.
\ee
The dual vector $\dvac$ is defined similarly
\be{0dvac} 
\dvac A(\lambda)=a(\lambda)\dvac,\quad 
\dvac D(\lambda)= d(\lambda)\dvac,\quad \dvac 
B(\lambda)=0.
\ee
The explicit form of functions  $a(\lambda)$ and $d(\lambda)$, entering 
Eqs. \eq{0vac}, \eq{0dvac}, depends on specific model.

The $R$-matrix, defining the commutation relations of  the monodromy 
matrix entries \eq{0RTT-}, can be written in the form:  
\be{0Rm}
R(\lambda,\mu)=\left(
\begin{array}{cccc}
f(\mu,\lambda) &0&0&0\\
0&g(\mu,\lambda)&1&0\\
0&1&g(\mu,\lambda)&0\\
0&0&0&f(\mu,\lambda)
\end{array}\right).
\ee
For the models QNLS and $XXX$ Heisenberg chain
the $R$-matrix is rational:
\be{0fg} 
f(\mu,\lambda)=\frac{\mu-\lambda+ic}{\mu-\lambda},\qquad
g(\mu,\lambda)=\frac{ic}{\mu-\lambda}.
\ee
Here $c$ is a constant. Below we shall use also two new functions 
\be{0ht}
\begin{array}{rcl}{\dis 
h(\mu,\lambda)}&
{\dis \equiv\frac{f(\mu,\lambda)}{g(\mu,\lambda)}}&
{\dis =\frac{\mu-\lambda+ic}{ic},}\non
{\dis 
t(\mu,\lambda)}&{\dis 
\equiv\frac{g(\mu,\lambda)}{h(\mu,\lambda)}}&
{\dis =-\frac{c^2}{(\mu-\lambda)(\mu-\lambda+ic)}.}
\end{array}
\ee

For the $XXZ$ chain the $R$-matrix has the same form 
\eq{0Rm}, however the functions $g(\lambda,\mu)$ and $f(\lambda,\mu)$
become trigonometric ones
\be{0fgtrig} 
f(\mu,\lambda)=\frac{\sinh(\mu-\lambda+i\gamma)}
{\sinh(\mu-\lambda)},\qquad
g(\mu,\lambda)=\frac{i\sin\gamma}{\sinh(\mu-\lambda)}.
\ee
Respectively the functions $h(\lambda,\mu)$ and $t(\lambda,\mu)$
should be replaced by their trigonometric analogs according to
the definition \eq{0ht}. The constant $\gamma$ is related to the
anisotropy of the model by $\cos\gamma=\Delta$.

The entries of the monodromy matrix \eq{0mon} act in a space,
consisting of states
\be{0ABA} 
|\Psi_N(\{\lambda\})\rangle=\prod_{j=1}^N 
B(\lambda_j)\vac,\qquad N=0,1,\dots
\ee
Here $\{\lambda\}$ are arbitrary complex parameters. Similarly dual 
states can be constructed by the operators $C(\lambda)$ 
\be{0dABA} 
\langle\Psi_N(\{\lambda\})|=\dvac\prod_{j=1}^N C(\lambda_j),
\qquad N=0,1,\dots 
\ee 
In what follows we shall refer to the states \eq{0ABA}, \eq{0dABA}
with arbitrary $\lambda$-s as `BA-vectors' (Bethe Ansatz vectors).

The transfermatrix $\tau(\lambda)=\mathop{\rm tr} T(\lambda)
=A(\lambda)+D(\lambda)$ generates the complete set of the conservation
laws of the model. The eigenstates of the transfermatrix have the
form \eq{0ABA}, \eq{0dABA}, however the parameters $\{\lambda\}$
are not arbitrary, but satisfy the system of Bethe equations
\be{0Betheeq} 
r(\lambda_j)\prod_{k=1\atop{k\ne j}}^N
\frac {f(\lambda_j,\lambda_k)} {f(\lambda_k,\lambda_j)}=1, \qquad 
\mbox{where}\qquad r(\lambda)=\frac{a(\lambda)}{d(\lambda)}.
\ee
The eigenvalues of the transfermatrix are
\be{0eigenval}
\begin{array}{l}
\dis \tau(\mu)|\Psi_N(\{\lambda\})\rangle=
\theta(\mu|\{\lambda\})|\Psi_N(\{\lambda\})\rangle,\non
\dis \theta(\mu|\{\lambda\})=
a(\mu)\prod_{m=1}^{N}f(\mu,\lambda_m)+
d(\mu)\prod_{m=1}^{N}f(\lambda_m,\mu).
\end{array}
\ee

Let us focus now at the models, which will be considered in the next
sections. For the QNLS the pseudovacuum \eq{0vac} coincides with
the Fock vacuum \eq{0Bosefield}. The coupling constant $c$ is equal
to the constant in \eq{0fg}, the functions $a(\lambda)$ and 
$d(\lambda)$ are
\be{0QNLSad}
a(\lambda)=\exp(-i\lambda L/2),\qquad
d(\lambda)=\exp(i\lambda L/2).
\ee

In the $XXZ$ chain with $M$ sites the pseudovacuum vectors
$\vac$ and $\dvac$ are
\be{0XXZvac}
\vac=\mathop{\otimes}\limits_{m=1}^M \uparrow_m,\qquad
\dvac=(\vac)^T.
\ee
For intermediate calculations the inhomogeneous $XXZ$ chain is
suitable. For this case the monodromy matrix is given by
\be{0XXZmoninhomo}
T(\lambda)=L_M(\lambda-\xi_M)\cdots L_1(\lambda-\xi_1),
\ee
where $\xi_m$ are arbitrary complex numbers, and
local $L$-operators are
\be{0XXZLoper}
L_m(\lambda)=i\left(
\begin{array}{cc}
\cosh\Bigl(\lambda-i\frac{\gamma}{2}\sigma_z^{(m)}\Bigr)  &
\sin\gamma\sigma_-^{(m)}  \\
\sin\gamma\sigma_+^{(m)}  &
\cosh\Bigl(\lambda+i\frac{\gamma}{2}\sigma_z^{(m)}\Bigr)  
\end{array}\right)
\ee
The functions $a(\lambda)$ and $d(\lambda)$ are equal to
\be{0XXZad}
a(\lambda)=\prod_{m=1}^{M}\cosh(\lambda-\xi_m-i\gamma/2),\qquad
d(\lambda)=\prod_{m=1}^{M}\cosh(\lambda-\xi_m+i\gamma/2).
\ee
In the limit $\xi_m\to0$ we turn back to the homogeneous
$XXZ$ chain.

Particular case $\Delta=1$ corresponds to $XXX$ Heisenberg
magnet. This model is described by rational $R$-matrix \eq{0Rm}
with $c=1$. The vacuum eigenvalues are
\be{0XXXad}
a(\lambda)=\prod_{m=1}^{M}(\lambda-\xi_m-i/2),\qquad
d(\lambda)=\prod_{m=1}^{M}(\lambda-\xi_m+i/2).
\ee

In the conclusion of this section we would like to mention that the
Hamiltonians, presented in the Introduction, are not of the most 
general form.  One can consider, for instance, Heisenberg magnets in 
external magnetic field. Similarly it is possible to consider QNLS 
with a chemical potential. In order to do this, one need to add to 
the Hamiltonians \eq{0HamQNLS}, \eq{0HamXXZ} the corresponding 
terms. However these additional terms do not play an essential role 
in the models in finite volume.  They become significant only in the 
thermodynamic limit, in particular they define the structure of the 
ground state and the thermodynamic properties of the models. Since  
the thermodynamic limit is not considered in this paper, therefore 
we restrict our selves with the Hamiltonians \eq{0HamQNLS} and 
\eq{0HamXXZ} only.

\section{Form factor of the local field in the QNLS}

The calculation of form factors in the framework of the Algebraic
Bethe Ansatz formally can be reduced to the calculation of scalar
products of a special type:
\be{1specscalpr}
S_N(\{\mu\},\{\lambda\})=\langle0|\left(\prod_{j=1}^N 
C(\mu_j)\right)\left(\prod_{j=1}^N B(\lambda_j)\right)|0\rangle.
\ee
In the last equation at least one of the states, say 
$\langle0|\prod_{j=1}^N C(\mu_j)$, is an eigenfunction of the 
transfermatrix. In other words the parameters $\mu_1,\dots,\mu_N$ 
satisfy the system of the Bethe equations \eq{0Betheeq}, while the 
parameters $\lambda_1,\dots,\lambda_N$ remain free, i.e.  
$\prod_{j=1}^NB(\lambda_j)$ is an arbitrary BA-vector.

Let us demonstrate the general idea of the method, which  will be 
used.  Consider a form factor of an operator $\cal O$ 
\be{1FFO} F_{\cal O}=
\langle0|\left(\prod_{j=1}^{N_\mu} C(\mu_j)\right) 
{\cal O}\left(\prod_{j=1}^{N_\nu} B(\nu_j)\right)|0\rangle.  
\ee 
Here both states $\langle0|\prod_{j=1}^{N_\mu} C(\mu_j)$ and 
$\prod_{j=1}^{N_\nu} B(\nu_j)|0\rangle$ are eigenstates of the
transfermatrix. The operator $\cal O$ acting on the vector
$\prod_{j=1}^{N_\nu} B(\nu_j)|0\rangle$ gives a new state, which
generally speaking is not an eigenstate of $\tau(\lambda)$.
However, if this new state can be presented as a linear combination
of BA-vectors with numerical coefficients $c_k$
\be{1actionO}
{\cal O}\left(\prod_{j=1}^{N_\nu} B(\nu_j)\right)|0\rangle
=\sum_{k}c_k\left(\prod_{j=1}^{N_k} B(\lambda_j^{(k)})
\right)|0\rangle,
\ee
then evidently the form factor $F_{\cal O}$ is equal
\be{1FFOscprod}
F_{\cal O}=\sum_{k}c_k
\langle0|\left(\prod_{j=1}^{N_\mu} C(\mu_j)\right)
\left(\prod_{j=1}^{N_k} B(\lambda_j^{(k)})
\right)|0\rangle.
\ee
Anyway the state $\langle0|\prod_{j=1}^{N_\mu} C(\mu_j)$ remains
the eigenstate of the transfermatrix. Thus, in order to find 
explicit expression for the form factor $\cal O$, one need to 
calculate the scalar products \eq{1specscalpr}. In order to do this, 
it is enough to use the algebra \eq{0RTT-} only. Moreover the 
specific form of the vacuum eigenvalues $a(\lambda)$ and
$d(\lambda)$, as well as the structure of the pseudovacuum, are
not essential. However, the straightforward calculation of the scalar 
products appears to be a complicated combinatorial problem. These
difficulties had been partly overcame in \cite{K0}, \cite{IK}.
In the paper \cite{K1} the determinant representation for the scalar
product of two arbitrary BA-vectors was obtained in terms of
auxiliary quantum operators---dual fields. The explicit determinant 
representation for the scalar products \eq{1specscalpr} was 
found in \cite{S1} (see also \cite{S2}):                                  
\ba{1detrepscalprod}
&&{\dis\hspace{-15mm}
S_N(\{\mu\},\{\lambda\})=
\left\{\prod_{a=1}^Nd(\mu_a)d(\lambda_a)\right\}
\left\{\prod_{a,b=1}^Nh(\mu_a,\lambda_b)\right\}}\non
&&{\dis\hspace{25mm}\times
\left\{\prod_{N\geq a >b \geq 1}g(\lambda_a,\lambda_b)
g(\mu_b,\mu_a)\right\}
\det\left(M_{jk}\right),}
\ea
where
\be{1Mjk}
M_{jk}=t(\mu_k,\lambda_j)-r(\lambda_j)t(\lambda_j,\mu_k)
\prod_{m=1}^N\frac{f(\lambda_j,\mu_m)}{f(\mu_m,\lambda_j)}.
\ee
Recall that here the parameters $\{\mu\}$ satisfy the Bethe 
equations \eq{0Betheeq}, while the parameters $\{\lambda\}$ are
free. Before applying this formula to the calculations of form factors
we would like to mention several features of the representation
\eq{1detrepscalprod}.

This representation is valid for arbitrary model, possessing rational
$R$-matrix \eq{0Rm} or trigonometric one.  The representation 
\eq{1detrepscalprod} also allows one to prove the orthogonality of 
the transfermatrix eigenstates directly. Indeed, the particular case 
of an BA-vector $\prod_{j=1}^{N} B(\lambda_j)|0\rangle$ is an 
eigenstate of $\tau(\lambda)$. Therefore, one can demand the 
parameters $\{\lambda\}$ to satisfy the Bethe equations 
\eq{0Betheeq}, but to be different from the parameters $\{\mu\}$. In 
this case we obtain the scalar product of two different eigenstates, 
which must be equal to zero. It is easy to check, that for the such 
choice of $\{\lambda\}$ the determinant $\det M$ in 
\eq{1detrepscalprod} does vanishes. Indeed, the matrix \eq{1Mjk} 
turns into                
\be{1Mjk2}
\tilde M_{jk}=t(\mu_k,\lambda_j)+V_jt(\lambda_j,\mu_k)
\ee
where
\be{1V}
V_j=\prod_{m=1}^N
\frac{h(\lambda_j,\mu_m)h(\lambda_m,\lambda_j)}
{h(\mu_m,\lambda_j)h(\lambda_j,\lambda_m)}.
\ee
This matrix has the eigenvector $\xi$ with zero eigenvalue:
\be{1eigenvector}
\sum_{k=1}^{N}\tilde M_{jk}\xi_k=0,\qquad
\xi_k=\frac{\prod\limits_{m=1\atop{m\ne k}}^{N}g(\mu_k,\mu_m)}
{\prod\limits_{m=1}^{N}g(\mu_k,\lambda_m)}.
\ee
The proof of the \eq{1eigenvector} is given in Appendix 1.
Hence $\det\tilde M=0$.

Now let us apply the representation \eq{1detrepscalprod} to the
calculation of specific form factors. Namely, consider the
form factor of the local field $\Psi(x,t)$ in the QNLS
\be{1fffieldxt}
F_{\Psi}(x,t)=\langle0|\left(\prod_{j=1}^{N} C(\mu_j)\right)
\Psi(x,t)\left(\prod_{j=1}^{N+1} B(\lambda_j)\right)|0\rangle.
\ee
The dependency on time $t$ and distance $x$ evidently separated
\be{1fffield}
F_{\Psi}(x,t)=
\exp\left\{\sum_{k=1}^{N}\left(it\mu_k^2-ix\mu_k\right)-
\sum_{j=1}^{N+1}\left(it\lambda_j^2-ix\lambda_j\right)\right\}
F_{\Psi}(0,0).
\ee
The action of the operator $\Psi(0,0)$ on the BA-vector had been
found in \cite{IKR}:   
\be{1actionPsi}
\Psi(0,0)\prod_{j=1}^{N+1}B(\lambda_j)|0\rangle=
-i\sqrt{c}\sum_{\ell=1}^{N+1}a(\lambda_\ell)
\left(\prod_{m=1\atop{m\ne \ell}}^{N+1}f(\lambda_\ell, \lambda_m)\right)
\prod_{m=1\atop{m\ne\ell}}^{N+1}B(\lambda_m)|0\rangle.
\ee
In spite of the parameters $\lambda_1,\dots,\lambda_{N+1}$ satisfy
the Bethe equations 
\be{1Bethe}
r(\lambda_j)=\prod_{p=1}^{N+1} \frac{h(\lambda_p,\lambda_j)}{
h(\lambda_j,\lambda_p)},
\ee
the states entering the r.h.s. of \eq{1actionPsi} are not
eigenstates of the transfermatrix, since each of them depends
only on $N$ parameters $\lambda$. Nevertheless all these states
are BA-vectors, therefore we can apply \eq{1detrepscalprod} for
calculation of the scalar products
\be{1scalprff}
S_N^{(\ell)}(\{\mu\},\{\lambda\})=\langle0|\left(\prod_{j=1}^N 
C(\mu_j)\right)\left(\prod_{j=1\atop{j\ne\ell}}^{N+1} 
B(\lambda_j)\right)|0\rangle,
\ee 
since the state $\langle0|\prod_{j=1}^{N}C(\mu_j)$ remains an
eigenstate. After simple algebra \cite{KKS1} we obtain for the form 
factor $F_\Psi(0,0)$:  
\ba{1ffinterm}                 
&&{\dis\hspace{-10mm} F_\Psi(0,0)=
	-i\sqrt{c}
	\prod_{a>b}^Ng(\mu_a, \mu_b)
	\prod_{a>b}^{N+1}g(\lambda_b, \lambda_a)
	\prod_{a,b=1}^{N+1}h(\lambda_a,\lambda_b)
}\non
&&{\dis\hspace{10mm}
	\times
	\prod_{a=1}^{N}d(\mu_a)
	\prod_{b=1}^{N+1}d(\lambda_b)
	\left(
	\sum_{\ell=1}^{N+1}(-1)^{\ell+1}
	{\det} S^{(\ell)}_{jk}
	\right).
}
\ea
Here we introduce new matrix $S$:
\be{1S}
S_{jk}=
t(\mu_k,\lambda_j)
\frac{\prod\limits_{m=1}^N h(\mu_m,\lambda_j)}{
\prod\limits_{p=1}^{N+1}h(\lambda_p,\lambda_j)}
-t(\lambda_j,\mu_k)
\frac{\prod\limits_{m=1}^N h(\lambda_j,\mu_m)}{
\prod\limits_{p=1}^{N+1}h(\lambda_j,\lambda_p)},
\qquad
j,k=1,\dots,N.
\ee
The matrix $S^{(\ell)}$ can be obtained from the matrix $S$
via the replacement of the $\ell$-th row by $S_{N+1,k}$:
\be{1Sell}
S_{jk}^{(\ell)}=\left\{
\begin{array}{ll}
S_{jk},&\qquad j\ne\ell,\\
S_{N+1,k},&\qquad j=\ell.
\end{array}
\right.
\ee

The sum with respect to $\ell$ in \eq{1S} gives a single 
determinant: 
\be{1singdet}
\sum_{\ell=1}^{N+1}(-1)^{\ell+1}{\det} S^{(\ell)}_{jk}
={\det}(S_{jk}-S_{N+1,k}).
\ee
In order to prove \eq{1singdet} one can use the Laplace formula
for determinant of sum of two matrices, and to take into account 
that $S_{N+1,k}$ is the first-rank matrix. Thus we arrive at
the determinant representation for the form factor of the
local field ib QNLS \cite{KKS1}:
\ba{1ffdetrep}                 
&&{\dis\hspace{-10mm} F_\Psi(0,0)=
	-i\sqrt{c}
	\prod_{a>b}^Ng(\mu_a, \mu_b)
	\prod_{a>b}^{N+1}g(\lambda_b, \lambda_a)
	\prod_{a,b=1}^{N+1}h(\lambda_a,\lambda_b)
}\non
&&{\dis\hspace{10mm}
	\times
	\prod_{a=1}^{N}d(\mu_a)
	\prod_{b=1}^{N+1}d(\lambda_b)
	{\det}(S_{jk}-S_{N+1,k}).
}
\ea
Here $S_{jk}$ is given by \eq{1S}, and $S_{N+1,k}$ can be obtained
from $S_{jk}$ via the replacement $\lambda_j\to\lambda_{N+1}$.

In the conclusion of this section we would like to emphasize once
more that the most serious problem of the described method is
to embed the operator $\cal O$, whose form factor wanted to be found,
into the algebra \eq{0RTT-}. However, after the commutation relations
between $\cal O$ and the monodromy matrix are defined, then the
problem can be solved via standard methods of the Algebraic Bethe Ansatz.

\section{Other representations for form factors}

As we have mentioned already, the method used for calculations of 
form factors of relativistic models is essentially different from 
the method, presented in the previous section. 
Nevertheless a series of the results, obtained in this domain for
the quantum field theory models, also can be applied for the models in
finite volume. Thus, for instant, the  form factors
of $O(3)$-invariant $\sigma$-model unsuspectingly had been found
closely related to the form factors of the QNLS \cite{KiSmO3}.
In the present section we consider several examples.
				       
The approach, developed in papers \cite{K0,IK}, allows to
express form factors of the QNLS in terms of some rational functions 
$\Sigma^\alpha$ (originally they were denoted as 
$\sigma^\alpha$), which depend on $N_\mu$ 
arguments $\{\mu_a\}$ and $N_\lambda$ arguments $\{\lambda_b\}$. The 
only singularities of these functions are simple poles at 
$\lambda_j=\mu_k$, and the residues in these points can be
expressed in terms of $\Sigma^\alpha$, depending on 
the arguments $\{\mu_{a\ne k}\}$ and $\{\lambda_{b\ne j}\}$. This 
property allows one to construct $\Sigma^\alpha$ via a recurrence,
however during a long time this recurrence have not been solved
in general form.

On the other hand the same functions $\Sigma^\alpha$ appeared to be 
useful for the analysis of the form factors of $O(3)$-invariant 
$\sigma$-model.  The series of the explicit determinant 
representations for $\Sigma^\alpha$, depending on elementary 
symmetric polynomials $\sigma^{(n)}_k$, were found in \cite{KiSm}:  
\be{2elsimpol} 
\sigma^{(n)}_k(x_1,\dots,x_n)=\frac{1}{(n-k)!}
\left.\frac{\partial^{(n-k)}}{\partial x^{(n-k)}}
\prod_{m=1}^{n}(x+x_m)\right|_{x=0}.
\ee

The explicit solution for the functions $\Sigma^\alpha$ has
the form:
\be{2detsigma}
\Sigma^\alpha(\{\mu\},\{\lambda\})=
\frac{\det\Omega}{\Delta(\mu)\Delta(\lambda)
\prod_{a,b}(\mu_a-\lambda_b)}.
\ee
Here $(N_\mu+N_\lambda)\times(N_\mu+N_\lambda)$ matrix $\Omega$ 
has the following entries:
\ba{2blokM}
&&{\dis\hspace{-1cm}
\Omega_{j,k}\equiv M_{jk}=
\sigma_j^{(N_\lambda+N_\mu-1)}\left(\{\mu_{a\ne k}-\frac{ic}2\},
\{\lambda_a+\frac{ic}2\}\right)}\non
&&{\dis\hspace{1cm}
-e^\alpha\sigma_j^{(N_\lambda+N_\mu-1)}
\left(\{\mu_{a\ne k}+\frac{ic}2\},
\{\lambda_a-\frac{ic}2\}\right),\qquad k=1,\dots,N_\mu,}
\ea
\ba{2blokL}
&&{\dis\hspace{-1cm}
\Omega_{j,k+N_\mu}\equiv \Lambda_{jk}=
\sigma_j^{(N_\lambda+N_\mu-1)}\left(\{\mu_a+\frac{ic}2\},
\{\lambda_{a\ne k} -\frac{ic}2\}\right)}\non
&&{\dis\hspace{1cm}
-\sigma_j^{(N_\lambda+N_\mu-1)}\left(\{\mu_a-\frac{ic}2\},
\{\lambda_{a\ne k}+\frac{ic}2\}\right),\qquad k=1,\dots,N_\lambda.}
\ea
Index $j$ runs through $0,1,\dots,N_\mu+N_\lambda-1$.\ By 
$\Delta(\cdot)$ we denote the Wandermonde determinants.

Form factors of the QNLS can be expressed in terms of the 
functions $\Sigma^\alpha$. In particular the form factor of the 
local field $F_\Psi(0,0)$, considered in the previous section,
is proportional to the function $\Sigma^0$ with $N_\mu=
N_\lambda-1=N$:
\be{2fffield}
F_\Psi(0,0)=-i\sqrt{c}
	\prod_{a=1}^{N}d(\mu_a)
	\prod_{b=1}^{N+1}d(\lambda_b)
\left.\Sigma^\alpha(\{\mu\}_N,\{\lambda\}_{N+1})
\vphantom{2^2}\right|_{\alpha=0}.
\ee
It is wort mentioning, that in this case $\Omega$
is $(2N+1)\times(2N+1)$  matrix. On the other hand, as we have
seen above, the form factor of the local field is proportional to the
determinant of the $N\times N$ matrix \eq{1Mjk}. There exists a nice 
method to reduce the determinant of the matrix $\Omega$ \eq{2blokM},
\eq{2blokL} to the determinant of \eq{1Mjk}. We shall demonstrate 
this method for more simple form factor.

Consider an operator $Q_1$ of the number of particles in the interval
$[0,x]$. In the QNLS this operator is equal to
\be{2Q1}
Q_1=\int_0^x\Psi^\dagger(y)\Psi(y)\,dy.
\ee
The form factor of this operator was calculated in \cite{IK}
in terms of the function $\Sigma^\alpha$ with $N_\mu=N_\lambda=N$:
\ba{2ffQ1}
&&{\dis\hspace{-1cm}
F_{Q_1}\equiv\langle0|\left(\prod_{j=1}^{N}C(\mu_j)\right)
Q_1\left(\prod_{j=1}^{N}B(\lambda_j)\right)|0\rangle}\non
&&{\dis\hspace{1cm}
=\left[\exp\left\{ix\sum_{a=1}^{N}
(\lambda_a-\mu_a)\right\}-1\right]
\left.\frac{\partial}{\partial\alpha}\Sigma^\alpha
(\{\mu\}_N,\{\lambda\}_N)\right|_{\alpha=0}.}
\ea
Let us show, how one can reduce the determinant of
$2N\times2N$ matrix \eq{2detsigma} to a determinant of $N\times N$ 
matrix.  
\begin{thm} 
The function $\Sigma^\alpha(\{\mu\}_N,\{\lambda\}_N)$ is proportional
to the determinant of $N\times N$ matrix
\ba{2Moasigma}
&&{\dis
\Sigma^\alpha(\{\mu\}_N,\{\lambda\}_N)=\prod_{a>b}^{N}
g(\lambda_j,\lambda_k)g(\mu_k,\mu_j)\prod_{a,b=1}^{N}
h(\lambda_a,\mu_b)}\non
&&{\dis\hspace{3cm}
\times\det\left(e^\alpha V_jt(\mu_k,\lambda_j)
+t(\lambda_j,\mu_k)\right),}
\ea
where
\be{2V}
V_j=\prod_{m=1}^{N}
\frac{h(\mu_m,\lambda_j)h(\lambda_j,\lambda_m)}
{h(\lambda_j,\mu_m)h(\lambda_m,\lambda_j)}.
\ee
(compare with \eq{1V}).
\end{thm}
{\sl Proof.}~~The determinant representation for the function
$\Sigma^\alpha(\{\mu\}_N,\{\lambda\}_N)$ is given by 
\eq{2detsigma}--\eq{2blokL}  with $N_\mu=N_\lambda=N$. Let us 
multiply the matrix $\Omega$ from the left by matrix $U$
\be{2blokP}
U_{j,k}\equiv P_{jk}=p^{2N-k-1}_j, \qquad j=1,\dots,N,
\ee
\be{2blokQ}
U_{j+N,k}\equiv Q_{jk}=q^{2N-k-1}_j, \qquad j=1,\dots,N.
\ee
Index $k$ runs through the set $0,1,\dots,2N-1$, parameters
$p_j$ and $q_j$ are some complex numbers, which will be fixed later.
Obviously
\be{2detOmega}
\det\Omega=\frac{\det(U\Omega)}{\det U},
\ee
and
\be{2detU}
\det U=\Delta(p)\Delta(q)\prod_{a,b=1}^{N}(p_a-q_b).
\ee
The product $U\Omega$ can be written as $2\times2$ block-matrix
\be{2blokmatrix}
U\Omega=\left(
\begin{array}{cc}
PM&P\Lambda\\
QM&Q\Lambda
\end{array}\right),
\ee
where each of these blocks is  $N\times N$ matrix. Using the 
representation for elementary symmetric polynomials \eq{2elsimpol}, 
one can easily find the entries of these blocks, for example:
\ba{2blockPM}
&&{\dis\hspace{-2cm}
(PM)_{jk}(\{p\},\{\mu\},\{\lambda\})=
\prod_{a=1\atop{a\ne k}}^{N}\left(p_j+\mu_a-\frac{ic}{2}\right)
\prod_{a=1}^{N}\left(p_j+\lambda_a+\frac{ic}{2}\right)}\nona{20}
&&{\dis\hspace{1cm}-e^\alpha  
\prod_{a=1\atop{a\ne k}}^{N}\left(p_j+\mu_a+\frac{ic}{2}\right)
\prod_{a=1}^{N}\left(p_j+\lambda_a-\frac{ic}{2}\right),}
\ea
Other blocks have similar form. Namely, replacing $p_j$
by $q_j$ in \eq{2blockPM} we obtain the block $QM$. As for
the right blocks, they can be obtained from the left ones, in
which one should replace $\{\mu\}$ by $\{\lambda\}$ and put 
$\alpha=0$.  Thus we obtain the representation for the function 
$\Sigma^\alpha(\{\mu\}_N,\{\lambda\}_N)$, containing $2N$ arbitrary
complex parameters $p_j$ and $q_j$, but does not depending on their
specific choice. 

Using well known formula for a determinant of a block-matrix
\be{2detblock}
\det\left(\begin{array}{cc}
A&B\\C&D\end{array}\right)
=\det A\cdot\det(D-CA^{-1}B),
\ee
one can reduce $\det(U\Omega)$ to the determinants of $N\times N$
matrices. However, in order to simplify the calculations, we 
shall fix the parameters $p_j$ and $q_j$. Let, for example, 
$p_j=\lambda_j-ic/2$ and $q_j=\lambda_j+ic/2$. Then we have for
two left blocks:
\be{2blokPMn}
(PM)_{jk}=-e^\alpha  
\prod_{a=1\atop{a\ne k}}^{N}(\mu_a-\lambda_j)
\prod_{a=1}^{N}(\lambda_{aj}-ic),
\ee
\be{2blokQMn}
(QM)_{jk}=  
\prod_{a=1\atop{a\ne k}}^{N}(\mu_a-\lambda_j)
\prod_{a=1}^{N}(\lambda_{aj}+ic),
\ee
Here $\lambda_{aj}=\lambda_a-\lambda_j$. Observe that
\be{2observe}
(QM)_{jk}= -u_j(PM)_{jk},\qquad\mbox{where}\qquad
u_j=e^{-\alpha}\prod_{a=1}^{N}\left(\frac{\lambda_{aj}+ic}
{\lambda_{aj}-ic}\right).
\ee
Hence, due to \eq{2detblock} we have
\be{2proddet}
\det(U\Omega)=\det(PM)_{jk}\cdot\det\Bigl(
(Q\Lambda)_{jk}+u_j(P\Lambda)_{jk}\Bigr).
\ee
After simple algebra we arrive at
\be{2elemtrans}
\det(U\Omega)=W\det G
\ee
where
\be{2W}
W=\prod_{a,b=1}^{N}\left[(\mu_a-\lambda_b)(\lambda_{ab}-ic)
(\lambda_{ab}+ic)\right]\prod_{a>b}^{N}[\mu_{ab}\lambda_{ba}],
\ee
and
\ba{2Gjk}
&&{\dis\hspace{-1cm}
G_{jk}=
\delta_{jk}\frac{\prod\limits_{a=1\atop{a\ne j}}^{N}\lambda_{aj}}
{\prod\limits_{a=1}^{N}(\mu_a-\lambda_j)}
\left\{e^\alpha
\prod_{a=1}^{N}\left(\frac{\mu_a-\lambda_j+ic}
{\lambda_{aj}+ic}\right)-
\prod_{a=1}^{N}\left(\frac{\mu_a-\lambda_j-ic}
{\lambda_{aj}-ic}\right) \right\}}\non
&&{\dis\hspace{5cm} 
+\frac{e^\alpha}{\lambda_{jk}-ic}
-\frac{1}{\lambda_{jk}+ic}.}
\ea
In order to reduce $\det G$ to the determinant \eq{2V}, one need to make
one more step:
\be{2XB+D}
\det G=\frac{\det(G\Gamma)}{\det \Gamma},
\ee
where
\be{2defX}
\Gamma_{jk}=\frac{1}{\lambda_j-\mu_k}
\frac{\prod\limits_{a=1}^{N}(\lambda_j-\mu_a)}
{\prod\limits_{a=1\atop{a\ne j}}^{N}\lambda_{ja}},
\quad
\mbox{and}    \quad
\det \Gamma=\prod_{a>b}^{N}\frac{\mu_{ba}}{\lambda_{ba}}.
\ee
The product $G\Gamma$ can be computed explicitly via the method,
describing in the Appendix 1. The resulting representation  for 
the form factor $F_{Q_1}$ exactly coincides with \eq{2Moasigma}
\eq{2V}. 

One can use the same idea in order to reduce the determinant
of $(2N+1)\times(2N+1)$ matrix \eq{2fffield} to the determinant
of the $N\times N$ matrix \eq{1Mjk}. A small complication
is caused by the fact that in this case the blocks in the
matrix \eq{2blokmatrix} have different dimension, for example,
the block $PM$ is $N\times N$ matrix, while the block
$QM$ is $(N+1)\times N$ matrix etc. This leads to the appearance
of the term $S_{N+1,k}$ in \eq{1ffdetrep}.

Thus we have demonstrated different representations for form
factors and proved their equivalency. The representations in terms
of elementary symmetric polynomials are quite characteristic for the
form factors of the relativistic models (see, for instance,
\cite{KM}). At the same time the representations \eq{1ffdetrep},
\eq{2Moasigma} look more natural, since they depend only on the 
functions, entering the $R$-matrix of the model. These representations
also are suitable for the computation of the correlation functions.

\section{Form factors of Heisenberg chains}

As we have seen already, the calculation of the form factors
is based on the explicit determinant representation 
\eq{1detrepscalprod}, \eq{1Mjk} for the scalar products of the 
special type \eq{1specscalpr}. However the calculation it self of the 
scalar products in the framework of the Algebraic Bethe Ansatz deals 
with rather complicated combinatorial problems. The method based on 
the explicit representations for the functions $\Sigma^\alpha$ in 
terms of elementary symmetric polynomials also makes use of solving 
of complicated recurrence. In the paper \cite{M1} a new 
approach was proposed, which allows to simplify significantly the 
calculation of scalar products and form factors.  There a factorizing 
Drinfel'd twist $F$ was constructed. The new basis generated by the 
twist $F$ is extremely suitable to study the structure of BA-vectors.  
and for representation of local  operators 
in terms of the entries of the monodromy matrix. In the paper 
\cite{M2} this method was used for calculation of the form factors in 
Heisenberg chains. We would like to present here the main results of 
the paper \cite{M2}.

Consider inhomogeneous $XXZ$ ($XXX$) model \eq{0XXZmoninhomo}:
\be{3monodromy}
T(\lambda)=L_M(\lambda-\xi_M)\cdots L_1(\lambda-\xi_1).
\ee
Then the local spin operators $\sigma_a^{(m)}$ can be presented in 
terms of the monodromy matrix entries as 
\be{3sigma}
\begin{array}{l}
\dis \sigma_-^{(m)}=\left(\prod_{a=1}^{m-1}\tau(\xi_a)\right)
B(\xi_m)\left(\prod_{a=m+1}^{M}\tau(\xi_a)\right),\non
\dis \sigma_+^{(m)}=\left(\prod_{a=1}^{m-1}\tau(\xi_a)\right)
C(\xi_m)\left(\prod_{a=m+1}^{M}\tau(\xi_a)\right),\non
\dis \sigma_z^{(m)}=\left(\prod_{a=1}^{m-1}\tau(\xi_a)\right)
(A(\xi_m)-D(\xi_m))\left(\prod_{a=m+1}^{M}\tau(\xi_a)\right).
\end{array}
\ee
Here $\tau(\lambda)=A(\lambda)+D(\lambda)$ is the transfermatrix.

The representation \eq{3sigma} allows to find the form factors of
local spin operators of $XXZ$ ($XXX$) chains. Consider, for example,
form factor $F_-^{(m)}$ of the operator ${\sigma_-^{(m)}}$:

\be{3ffsigmam}
F_-^{(m)}=\langle0|\left(\prod_{j=1}^{N+1}C(\mu_j)\right)
\sigma_-^{(m)}\left(\prod_{j=1}^{N}B(\lambda_j)\right)|0\rangle.
\ee
Both states in \eq{3ffsigmam} are eigenstates of the transfermatrix,
hence substituting here \eq{3sigma} we obtain
\be{3ffsigmam1}
F_-^{(m)}=\prod_{a=1}^{m-1}\theta(\xi_a|\{\mu\})
\prod_{a=m+1}^{M}\theta(\xi_a|\{\lambda\})
\cdot\langle0|\left(\prod_{j=1}^{N+1}C(\mu_j)\right)
B(\xi_m)\left(\prod_{j=1}^{N}B(\lambda_j)\right)|0\rangle. 
\ee
Here $\theta$ is eigenvalue of the transfermatrix \eq{0eigenval}.
Thus we again have reduced the form factor to the scalar product
of eigenstate by BA-vector. It is enough now to use the 
representation \eq{1detrepscalprod}, which gives us the 
representation for the form factor $F_{\sigma_-^{(m)}}$ in terms
of a determinant of $(N+1)\times(N+1)$ matrix.

Let us present the final result in the form given in \cite{M2}. 
Introduce $(N+1)\times(N+1)$ matrix 
\be{3T} 
T_{jk}(\{\mu\}_{N+1}|\{\nu\}_{N+1})=
\frac{\partial}{\partial\nu_k}\theta(\mu_j|
\{\nu\}_{N+1}).
\ee
Then the form factor $F_-^{(m)}$ is proportional to the ratio
of two determinants
\be{3ff-res}
F_-^{(m)}=
\frac{\prod_{a=1}^{N+1}\prod_{b=1}^{m-1}f(\mu_a,\xi_b)}
{\prod_{a=1}^{N}\prod_{b=1}^{m}f(\lambda_a,\xi_b)}\cdot
\frac{\det T_{jk}(\{\mu\}|\{\xi_m,\lambda_1,\dots,\lambda_N\})}
{\det K_{jk}(\{\mu\}|\{\xi_m,\lambda_1,\dots,\lambda_N\})}.
\ee
Here $K_{jk}$ is the Cauchy matrix 
\be{3Cauchi}
K_{jk}(\{\mu\}|\{\nu\})=\varphi^{-1}(\mu_j-\nu_k),
\ee
where $\varphi(\lambda)=\lambda$ for $XXX$ chain and 
$\varphi(\lambda)=\sinh\lambda$ for $XXZ$ chain. Actually the function
$\varphi^{-1}(\lambda)$ coincides with the function $g(\lambda)$
up to constant factor. The representation similar to \eq{3ff-res}
also exists for the scalar products \eq{1detrepscalprod}.

\section{Conclusion}
We considered form factors of exactly solvable (completely 
integrable) models in the finite volume. We demonstrated that the form 
factors can be represented as determinants of matrices. The dimension 
of the matrix is related to the number of particles in the 
corresponding state.  In our earlier publications we showed that this 
representation is useful for the theory of correlation functions.  The 
contributions of all form factors can be taken into account and the 
correlation function also can be represented as a determinant.

\appendix

\section{Zero eigenvector} 
Consider for simplicity the case of the rational $R$-matrix.
Let $N\times N$ matrix \eq{1Mjk2} is given
\be{AMjk2} 
\tilde M_{jk}=t(\mu_k,\lambda_j)+V_jt(\lambda_j,\mu_k).
\ee
Here
\be{AV}
V_j=\prod_{m=1}^N
\frac{(\lambda_j-\mu_m+ic)(\lambda_m-\lambda_j+ic)}
{(\mu_m-\lambda_j+ic)(\lambda_j-\lambda_m+ic)}.
\ee
Recall also that $t(\lambda,\mu)=(ic)^2/(\lambda-\mu)
(\lambda-\mu+ic)$. Let us prove that 
\be{eigenvector}
\sum_{k=1}^{N}\tilde 
M_{jk}\xi_k=0, \qquad\mbox{where}\qquad
\xi_k=\frac{\prod\limits_{m=1}^{N}(\mu_k-\lambda_m)}
{\prod\limits_{m=1\atop{m\ne k}}^{N}(\mu_k-\mu_m)}.
\ee
We have
\be{subst}
\sum_{k=1}^{N}\tilde M_{jk}\xi_k=(ic)^2
\biggl(G_+ +V_jG_-\biggr),
\ee
where
\be{G1}
G_\pm=\sum_{k=1}^{N}\frac{1}
{(\mu_k-\lambda_j)(\mu_k-\lambda_j\pm ic)}
\cdot\frac{\prod\limits_{m=1}^{N}(\mu_k-\lambda_m)}
{\prod\limits_{m=1\atop{m\ne k}}^{N}(\mu_k-\mu_m)},
\ee
In order to find $G_\pm$ consider auxiliary integral
\be{I1}
I_\pm=\frac{1}{2\pi i}\int\limits_{|z|=R\to\infty}
\frac{dz}{(z-\lambda_j)(z-\lambda_j\pm ic)}
\prod\limits_{m=1}^{N}\frac{(z-\lambda_m)}{(z-\mu_m)}.
\ee
The integral is taken with respect to the closed contour around 
infinity (in other words all the poles of the integrand lie within 
the contour). Obviously the integral $I_\pm$ is equal to 
the residue at infinity, which in turn is equal to zero,
hence $I_\pm=0$.  On the other hand the value of the 
integral $I_\pm$ is equal to the sum of the residues in the poles 
$z=\mu_k,\quad k=1,\dots,N$ and $z=\lambda_j-ic$. Thus, we 
obtain 
\be{I12}
I_\pm=G_\pm\mp\frac{1}{ic}\prod_{m=1}^{N}
\left(\frac{\lambda_m-\lambda_j\pm ic}{\mu_m-\lambda_j\pm ic}
\right).
\ee
The first term in the r.h.s. of \eq{I12} corresponds to
the residues at the points $z=\mu_k,\quad k=1,\dots,N$; the second
term corresponds to the residue at the point $z=\lambda_j-ic$. 
Hence
\be{G1r}
G_\pm=\pm\frac{1}{ic}\prod_{m=1}^{N}
\left(\frac{\lambda_m-\lambda_j\pm ic}{\mu_m-\lambda_j\pm ic}
\right).
\ee
Substituting $G_\pm$ into \eq{subst} and using explicit
expression \eq{AV} for $V_j$, we arrive at
\be{proof}
\sum_{k=1}^{N}\tilde M_{jk}\xi_k=0.
\ee

In the case of trigonometric $R$-matrix,  after the replacement
$\exp\{2\lambda\}\to\lambda$ and $\exp\{2\mu\}\to\mu$ the proof 
reduces to the rational case.


\begin{thebibliography}{99}
%
\bibitem{1}
L.~D.~Faddeev and E.~K.~Sklyanin, Dokl. Acad. Nauk SSSR
{\bf 243} (1978) 1430.
%
\bibitem{2}
E.~K.~Sklyanin, Dokl. Acad. Nauk SSSR
{\bf 244} (1978) 1337.
%
\bibitem{3}
L.~D.~Faddeev, E.~K.~Sklyanin,  and L.~A.~Takhtajan, Theor. Math. 
Phys. {\bf 40} (1979) 194.  
%
\bibitem{4}
F.~A.~Smirnov, Form factors in completely integrable models of 
quantum field theory (World Scientific, Singapore, 1992).
%
\bibitem{ZZ} A.~B.~Zamolodchikov and Al.~B.~Zamolodchikov,
Ann. Phys.~120 (1979) 253; A.~B.~Zamolodchikov, Advanced Studies in 
Pure Math.~19 (1989) 641; Int.~J.~Mod.~Phys.~A 3 (1988) 743.
%
\bibitem{AFZ} A.~E.~Arinshtein, V.~A.~Fatteev, and A.~B.~Zamolodchikov,
Phys.~Lett.~B 87 (1979) 3389.
%
\bibitem{BKW} B.~Berg, M.~Karowski and P.~Weisz, Phys.~Rev.~D 
19 (1979) 2477;
M.~Karowski and P.~Weisz, Nucl.~Phys.~B 139 (1978) 445; 
M.~Karowski, Phys.~Rep. 49 (1979) 229.
%
\bibitem{AlZ} Al.~B.~Zamolodchikov, Nucl.~Phys.~B 348 (1991) 619.
%
\bibitem{CM} J.~L.~Cardy, G.~Mussardo, Nucl.~Phys.~B 340 (1990) 387.
%
\bibitem{KM} A.~Koubek and G.~Mussardo, Phys.~Lett.~B 
311 (1993) 193.
%
\bibitem{FMS} A.~Fring, G.~Mussardo and P.~Simonetti, Nucl.~Phys.~B 
393 (1990) 413. 
%
\bibitem{ADM} C.~Ahn, G.~Delfino and G.~Mussardo, 
Phys.~Lett.~B 317 (1993) 573. 
%
\bibitem{L}  S.~Lukyanov, Phys.~Lett.~B 408 (1997) 192;
V.~Brazh\-ni\-kov and S.~Lukyanov, preprints RU-97-58,
CLNS 97/1488, hep-th/9707091.
%
\bibitem{N} J.~Balog, T.~Hauer and M.~R.~Niedermaier, 
Phys.~Lett.~B 386 (1996) 224; 
Nucl.~Phys.~B 440 (1995) 603.
%
\bibitem{IKR}
A.~G.~Izergin, V.~E.~Korepin and N.~Yu.~Reshetikin,
J. Phys. {\bf A20} (1987) 4799.
%
\bibitem{K0} V.~E.~Korepin,
Commun. Math. Phys. {\bf 86} (1982) 391.
%
\bibitem{IK}
A.~G.~Izergin and V.~E.~Korepin,
Commun. Math. Phys. {\bf 99} (1985) 271.
%
\bibitem{K1} V.~E.~Korepin,
Commun. Math. Phys. {\bf 113} (1987) 177.
%
\bibitem{S1} 
N.~A.~Slavnov, Theor. Math. Phys. {\bf 79}
(1989) 502.
%
\bibitem{S2} 
N.~A.~Slavnov, Zap. Nauchn. Semin. POMI {\bf 245}
(1997) 270.
%
\bibitem{KKS1}
T.~Kojima, V.~E.~Korepin and N.~A.~Slavnov, 
Commun. Math. Phys. {\bf 188} (1997)   657. 
%
\bibitem{KiSmO3}
A.~N.~Kirillov and F.~A.~Smirnov, Int. Journ. Mod. Phys. A
{\bf 3} (1988) 731
%
\bibitem{KiSm}
A.~N.~Kirillov and F.~A.~Smirnov,  J. Sov. Math.
{\bf 47} (1989) 2413.
%
\bibitem{M1} 
J.~M.~Maillet and J.~Sanchez de Santos, Drinfel'd twists
and Algebraic Bethe Ansatz, q-alg/9612012.
%
\bibitem{M2}
N.~A.~Kitanin, J.~M.~Maillet and V.~Terras, Form factors of the
$XXZ$ Heisenberg spin-$1/2$ finite chain, math-ph/9807020.

\end{thebibliography}
\end{document}